\documentclass[prl,aps,twocolumn,showpacs]{revtex4}
\usepackage{epsfig}
\usepackage{wasysym}
\usepackage{bm}
\usepackage{pstricks}

\newlength{\myL}{

\newcommand{\beq}{\begin{equation}}
\newcommand{\eeq}{\end{equation}}
\newcommand{\bea}{\begin{eqnarray}}
\newcommand{\eea}{\end{eqnarray}}

\begin{document}

\title{Supersolidity, entropy and frustration}
\author{S. R. Hassan$^{1}$, L. de Medici$^{2},$ and A.-M.S.~Tremblay$^{1}$}
\affiliation{$^{1}$Department de physique and RQMP, Universit\'{e} de Sherbrooke,
Sherbrooke, Qu\'{e}bec, Canada J1K 2R1}
\affiliation{$^{2}$Department of Physics and Center for Materials Theory, Rutgers
University, Piscataway, NJ 08854, USA.}

\begin{abstract}
We study the properties of $t-t^{\prime }-V$ model of hard-core bosons on
the triangular lattice that can be realized in optical lattices. By mapping
to the spin-1/2 XXZ model in a field, we determine the phase diagram of the $%
t-V$ model where the supersolid characterized by the ordering pattern $%
(x,x,-2x^{\prime })$ (\textquotedblleft ferrimagnetic\textquotedblright\ or
SS\ A) is a ground state for chemical potential $\mu >3V$. By turning on
either temperature or $t^{\prime }$ at half-filling $(\mu =3V)$, we find a
first order transition from SS A to the elusive supersolid characterized by
the $(x,-x,0)$ ordering pattern (\textquotedblleft
antiferromagnetic\textquotedblright\ or SS\ C). In addition, we find a large
region where a superfluid phase becomes a solid upon raising temperature at
fixed chemical potential. This is an analog of the Pomeranchuk effect driven
by the large entropic effects associated with geometric frustration on the
triangular lattice.
%We search for the possibility of the non-Landau-Ginzburg-Wilson(LGW) deconfined quantum
%criticality in transition from supersolid SSA to SSC phases. We do not find any example of
%such  quantum criticality among any of the interesting phase transitions in the solid.
\end{abstract}

\maketitle

Supersolidity is one of the most intriguing properties of matter. In that
state, matter can flow without viscosity, like in a superfluid, yet atoms
are located at regular positions: Translation and $U\left( 1\right) $
symmetry are broken simultaneously. It was originally proposed \cite%
{SupersolidOriginal} that this state could exist in ${}^{4}$He. While such a
supersolid state may have been observed, \cite{Kim_Chan} it is likely that
the relevant mechanism for $^{4}$He is disorder \cite{Phillips:2007}, not
zero point vacancies as first envisioned.

To observe supersolidity without disorder, one can load ultracold bosonic
atoms into optical lattices \cite{Oplattices}. Indeed, Bose-Einstein
condensation (BEC) of chromium atoms in an optical trapping potential \cite%
{Griesmaier} has already been observed, making it likely that supersolid
phases on such lattices can eventually be achieved. Temperature is clearly
an extremely relevant parameter for these experiments. \cite{Boninsegni}

One of the most promising lattices to observe supersolid phases is the
triangular lattice where supersolidity appears as a result of geometric
frustration, from a kind of order-by-disorder mechanism \cite{Troyer, Damle,
paramekanti, Boninsegni,Tompsett,Melko}. Supersolidity in other
two-dimensional lattice models has been predicted theoretically, but the
triangular lattice offers a particularly rich and interesting phase diagram
in a lattice that is simple to realize. For example, it has been proposed 
\cite{DVT} that second-neighbor hopping may induce the intriguing
particle-hole symmetric supersolid C phase, (so-called \textquotedblleft
antiferromagnetic\textquotedblright\ supersolid). It has been conjectured 
\cite{DVT} that the transition between supersolid C and other phases, such
as supersolid A (\textquotedblleft ferrimagnetic\textquotedblright\
supersolid), could occur through a critical point with emergent degrees of
freedom that cannot be described by the standard Landau theory. \cite%
{Senthil}

% why this method...
In this paper, we obtain detailed phase diagrams showing that a
particle-hole symmetric supersolid phase C can indeed be stabilized by both
next-nearest-neighbor hopping and by finite temperature effects. In
addition, the frustration associated with the triangular lattice amplifies
entropic effects, leading to a wide range of parameters where one can
observe superfluid-solid-liquid transitions as temperature is raised at
constant chemical potential. On the square lattice \cite{Schmid:2002}, this
sequence of transitions occurs in an extremely narrow range of chemical
potentials. This phenomenon is an analog of the Pomeranchuk effect in $^{3}$%
He where liquid (not superfluid)-solid-liquid transitions are observed by
increasing $T$ at fixed pressure.

%\section{Methdology and model}
\textit{Model: }We consider hard core bosons (infinite on-site repulsion) on
a triangular lattice, with both nearest neighbor (nn) hopping and repulsion (%
$t,V$) and next nearest neighbor (nnn) hopping ($t^{\prime }$) 
\begin{equation}
H=-\sum_{i,j}t_{ij}a_{i}^{\dag }a_{j}+h.c+V\sum_{<ij>}n_{i}n_{j}-\mu
\sum_{i}n_{i}  \label{tVm}
\end{equation}%
where each lattice site can be occupied by $0$ or $1$ boson ($n_{i}=0,1$) $%
n_{i}=a_{i}^{\dag }a_{i}$, and $\mu $ is the chemical potential. In the
above restricted Hilbert space, the model (\ref{tVm}) can be mapped to the $%
S=1/2$ XXZ model in a field ($h$) 
\begin{equation}
{\mathcal{H}}=V\sum_{<ij>}S_{i}^{z}S_{j}^{z}-%
\sum_{i,j}t_{ij}S_{i}^{-}S_{j}^{+}+h.c.-h\sum_{i}S_{i}^{z}\;,  \label{XXZ}
\end{equation}%
where $h=\mu -3V$. In this language, supersolid (SS) ordering corresponds to
spins having their $x-y$ component aligned ferromagnetically (superfluid
(SF)) along with their $z$-component also ordered but at non zero wave
vector inside the first Brillouin zone (solid (S)). A phase without ordering
but non-zero $z$-component and zero $x-y$ component corresponds to the
normal fluid (NF). Fully polarized up (down) spins corresponds to Full
(Empty) lattice. The order parameter for the solid (staggered magnetization
in spin language, staggered density in boson language) is defined with the
help of the three sublattice magnetizations ($S_{i}^{z}=n_{i}-1/2,i=1,2,3)$
as \cite{Richter} $M_{s}=\sqrt{3(({S_{1}^{z}})^{2}+({S_{2}^{z}})^{2}+({%
S_{3}^{z}})^{2}-S_{1}^{z}S_{2}^{z}-S_{1}^{z}S_{3}^{z}-S_{2}^{z}S_{3}^{z})}.$
It measures the solid order, i.e. a periodicity longer than that of the
underlying lattice.

% The triangular lattice can be divided into six sublattices, which we 
%denote 1,2,3,4,5,6 as shown in Fig.\ref{fig:lattice}. 

\textit{Method: }The Berezinskii-Kosterlitz-Thouless (BKT) transitions and
the SS-C phase (in the $t/V\rightarrow 0$ limit) \cite{Boninsegni} are
normally out of reach for simple mean-field theories. Using large enough
clusters however, Self-consistent Cluster Mean Field Theory (SCMFT) can
overcome some of these deficiencies. We argue that, while not perfectly
accurate, SCMFT \cite{Zhao:2007} is an extremely efficient way of exploring
vast uncharted territory in the phase diagram. More refined studies can then
improve the accuracy of phase boundaries in a second stage. We briefly
describe the method and then demonstrate its accuracy by comparing with
known results.

A cluster $^{\prime }1^{\prime }$ with a finite number of sites (shaded
region of the inset of Fig. 1) is embedded in the effective field of its
surroundings. In other words we consider the following cluster $\mathcal{C}$%
\emph{\ }spin Hamiltonian ${\mathcal{H}}_{s}$
\begin{equation}
{\mathcal{H}}_{s}=\sum_{i,j\epsilon \mathcal{C}}{\mathcal{H}}%
_{ij}+\sum_{i\epsilon \mathcal{C}}h_{i}^{-}S_{i}^{\dag
}+h.c.+\sum_{i\epsilon \mathcal{C}}h_{i}^{z}S_{i}^{z}-h\sum_{i\epsilon
\mathcal{C}}S_{i}^{z}  \label{clusterH}
\end{equation}%
where $h_{i}^{-}$ and $h_{i}^{z}$ are the effective fields of the
surroundings. ${\mathcal{H}}_{s}$ needs to be diagonalized with the
following self-consistency conditions:
\begin{equation}
h_{i}^{-/+}=\sum_{j}^{\prime
}t_{ij}<S_{j}^{-/+}>,\,\,h_{i}^{z}=\sum_{j}^{\prime }V_{ij}<S_{j}^{z}>
\label{eq:selfc_field}
\end{equation}%
where $j$ indicates neighbor of site $i$ and prime over $\Sigma $ indicates
that sites $j$ inside the cluster are excluded. Average values in Eq.(\ref%
{eq:selfc_field}) are obtained from ${\mathcal{H}}_{s}.$

\textit{Validity of the approach: }To assess the accuracy of SCMFT, we first
show that it reproduces quite accurately the phase diagram obtained from the
most reliable approaches. From now on, we discuss the results mostly in the
bosonic language. For the $t-V$ model, consider a cluster $^{\prime
}1^{\prime }$ of three sites shown as a shaded area in the inset of Fig. 1.
We measure $t,\mu $, and temperature $T$ in units of $V$, defining $%
\widetilde{t}=t/V$, $\widetilde{\mu }=\mu /V$, and $\tilde{T}=T/V$.

\begin{figure}[tbp]
\includegraphics[scale=0.3]{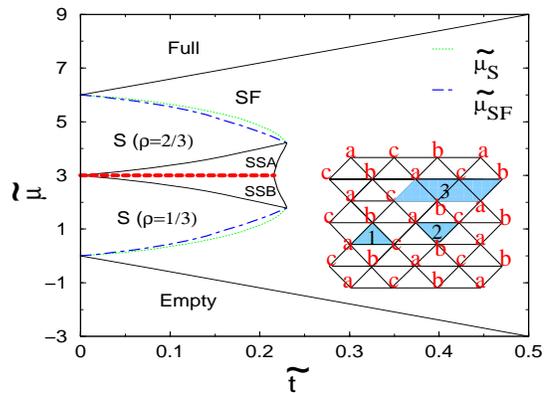}
\caption{(Color online) Zero-temperature phase diagram for the triangular
lattice. Second order phase transitions are denoted by solid lines, whereas $%
\widetilde{\protect\mu }_{S}$ and $\widetilde{\protect\mu }_{SF}$ are
spinodal lines as a function of inverse coupling strength $\widetilde{t}=t/V$%
. The thick lines dashed line at $\widetilde{\protect\mu }=3$ indicate first
order transition between SS A to SS B. Inset shows $\protect\sqrt{3}X\protect%
\sqrt{3}$ ordering of the solid and supersolid phases.}
\label{fig:Zerotph}
\end{figure}
We display the zero-temperature phase diagram in Fig.\ref{fig:Zerotph}. This
phase diagram is very close to the phase diagram obtained by Quantum Monte
Carlo (QMC) methods in Refs. \cite{Boninsegni,Troyer}. In the simplest
mean-field approach, \cite{Murthy} the supersolid region at $\widetilde{\mu }%
=3$ is much too large, extending to $(\widetilde{t})^{MF}=0.5$ compared with
$(\widetilde{t})^{QMC}=0.124$ in Ref. \cite{Troyer}. Here, we obtain $0.216$%
, closer to QMC. Also, in our approach, the maximum extent of the solid
region, $\widetilde{t}=0.22,$ is overestimated by only $10\%$ compared with
the QMC result $0.195$. In Fig. 1 we show the spinodals $\widetilde{\mu }%
_{S}(\widetilde{t})$ and $\widetilde{\mu }_{SF}(\widetilde{t})$ between
which metastable phases or coexistence of SF and S may occur.

In supersolid $A$ (SS A, $\widetilde{\mu }>3)$, the density on three
consecutive sites follows the \textquotedblleft
ferrimagnetic\textquotedblright\ ordering patterns $<n_{i}-\frac{1}{2}%
>=<S_{i}^{z}>=(x,x,-2x^{\prime }).$ In supersolid $B$ (\textit{SS B}, $%
\widetilde{\mu }<3$) the pattern is $(-x,-x,2x^{\prime })$ with $x\neq
x^{\prime }.$ This pattern is the same as that in Ref. \cite%
{Damle,Boninsegni}, in contrast with $x=x^{\prime }$ found in Ref. \cite%
{Troyer}. The density has a discontinuous jump at $\widetilde{\mu }=3,$
hence the \textit{SS\ A }- \textit{SS\ B} transition is first order. Larger
cluster size ($9$ sites) confirms this result. All these results (and more
below) validate the SCMFT approach to the hard-core boson problem. The
spinodal lines $\widetilde{\mu }\left( \widetilde{t}\right) $ for the
supersolid phases (not shown) have roughly a parabolic shape, closing at the
critical endpoints $(\widetilde{\mu }=3,\widetilde{t}=0)$ and $(\widetilde{%
\mu }=3,\widetilde{t}_{c}=0.216),$ the latter being the SS to SF transition.
The maximum size of the metastable region, $\widetilde{\mu }=\pm 3.01,$
occurs halfway between $\widetilde{t}=0$ and $\widetilde{t}_{c}.$

The main properties of the supersolid phases are summarized as follows at
the particle-hole symmetric point $\widetilde{\mu }=3$ (half-filling). When $%
\widetilde{t}$ approaches $0,$ the supersolid state is in close proximity to
the insulating states $\rho =2/3\;(\rho =1/3),$ therefore the jump in
density $\delta \rho $ is maximum in this region. The staggered density $%
M_{s}$ is also maximum there and vanishes continuously at the critical point
$\widetilde{t}_{c}=0.216$ after which only superfluidity survives$.$ The
superfluid density $\rho _{s}$ corresponds to the spin-stiffness in spin
language. It measures the energy cost to introduce a twist $\theta $ of the
direction of spin between every pair of neighboring rows. We use its
generalization to finite temperature following Ref. \cite{Zotos}. The SS A
to SF transition is a continuous quantum phase transition with a kink in $%
\rho _{s}$ at the transition point. The value of $\rho _{s}$ that we find
there ($0.18$) is within a few percent of the QMC results \cite{Troyer}.

%\begin{figure}
%\includegraphics[scale=0.3]{New_rho_vs_h_T_0.eps}
%\caption{ Density of hardcore bosons at zero temperature on the triangular
%lattice as a function of $h$ along lines of constant values of $\Delta$.
%Only densities $\rho\ge 1/2$ are shown since the phase diagram is symmetric
%around half-filling. Inset shows coexistance region of S and SF as a function
%of $\Delta$.}
%\label{fig:Zerotd}
%\end{figure}

%\begin{figure}

%    Since the phase diagram is symmeteric  when interchanging h (particles) with
%$-h$ (holes) we restrict our discussion  now on to $h\ge 0$ and plot the density
%$\rho$ as a function of magnetic field $h$ for cut at constant $\Delta=0.15 0.2,0.25$
%in Fig.(\ref{fig:Zerotd}). All these curves have been obtained by a feeding back
%previous value of h to a next value of h and where h is in increasing order.
%For $\Delta=0.1,0.2 $ we clearly observe  a plateau corresponding to the  $\rho=2/3(\rho=1/3)$ phases with broken translational symmetry.
%The approach to this plateau from $\rho<2/3(\rho>1/3)$ is continous,
%indicating second order phase transition, while for $\rho>2/3(\rho<1/3)$ we
%see a jump for the value of $\Delta=0.15, 0.2$ caused by a first order
%transition. In the inset we show the coexistance region of S and SF as
%function of $\Delta$.

\begin{figure}[tbp]
\includegraphics[scale=0.35]{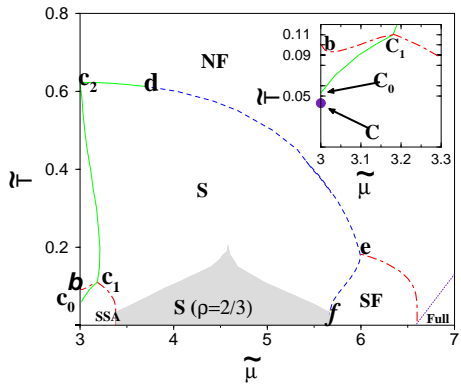}
\caption{(Color online) (a) $\widetilde{\protect\mu }$-$\tilde{T}$ Phase
diagram corresponding to a vertical line $\widetilde{t}=0.1$ in Fig. 1. The
inset shows the behavior near the particle-hole symmetric point $\widetilde{%
\protect\mu }=3.$ The two arrows indicate the region of metastability of
supersolid phases between $\tilde{T}_{c_{0}}$ and $\tilde{T}_{c}$. }
\label{fig:Finitetph}
\end{figure}
\begin{figure}[tbp]
\includegraphics[scale=0.3]{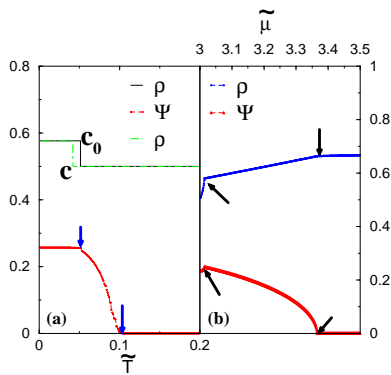}
\caption{(Color online) (a) Density $\left( \protect\rho \right) $
hysteresis and superfluid order parameter $\left( \Psi \right) $ as a
function of $\tilde{T}$ along the $\widetilde{\protect\mu }=3$ vertical line
of the phase diagram in Fig. 2. (b) Same quantities but this time as a
function of $\widetilde{\protect\mu }$ along the horizontal line $\tilde{T}%
=0.06$ in the phase diagram of Fig.2. }
\end{figure}

\textit{Finite temperature phase diagram at finite doping. }In Fig. 2 we
present the finite-temperature phase diagram along a vertical line $%
\widetilde{t}=0.1$ of Fig. 1. Because of particle-hole (Ising) symmetry it
is a sufficient to show $\widetilde{\mu }\geq 3$. Over a wide range of
chemical potentials at high temperatures, a first order S to NF phase
transition (dashed line) ends at a tricritical point $\mathbf{d}$ at about $%
\widetilde{\mu }=3.70$, where second order melting transition of the solid
begins. Point $\mathbf{e}$ at the other end of the first order line\textbf{\
}marks the beginning of a very interesting region at large $\widetilde{\mu }$%
. The first order transition bifurcates: to the right into a (BKT)
transition separating SF and NF and, to the left, into a first order
transition separating SF and S. Between point $\mathbf{e}$ and point $%
\mathbf{f,}$ we find the remarkable sequence of phases described in the
introduction: As we raise the temperature at fixed $\widetilde{\mu },$ one
encounters SF, S then NF. The superfluid solidifies as we increase
temperature because of an analog of the Pomeranchuk effect, the role of spin
entropy being played by hard-core boson occupation of optical lattice sites.
Solidification does not quench all the entropy. Let us come back to the BKT
transition to the right of point $\mathbf{e}$. One does expect the SF to NF
transition to be of this nature \cite{Schmid:2002}. Clearly, SCMFT cannot
accurately describe the topological BKT transition. Nevertheless, we take
the jump in superfluid density $\rho _{s}$ illustrated in the inset of Fig.
4,\ and the continuous vanishing of the order parameter $\Psi \equiv
\left\langle S_{x}\right\rangle =\left\langle \left( b+b^{\dagger }\right)
/2\right\rangle $ as very clear SCMFT signatures of the BKT transition.

Supersolid phases appear near the symmetric point $\widetilde{\mu }=3.$ For $%
\widetilde{\mu }\leq 3.38$ the solid freezes into various supersolid phases
with decreasing temperature. For example, at $\widetilde{\mu }=3.2$ the
staggered density $M_{s}$ and the density $\rho $ change continuously from S
to the finite $\widetilde{\mu }$ extension of SS\ A, but again there is a
jump in the $\rho _{s}$, as shown the inset of Fig. 4, so the transition is
of the BKT type.

The inset in Fig. 2 is a blow up of the region around the particle-hole
symmetric point $\widetilde{\mu }=3$ where supersolid phases appear. At $%
\widetilde{\mu }=3,$ raising $\tilde{T}$ from zero, we notice that the
ordering pattern of the solid changes from \textquotedblleft
ferrimagnetic\textquotedblright\ SS\ A $(x,x,-2x^{\prime })$ to
\textquotedblleft antiferromagnetic\textquotedblright\ SS\ C\ $(x,-x,0)$ at $%
\tilde{T}_{c_{o}}=0.053$, indicated by point $\mathbf{c_{0}}$ in the inset.
The SS\ A to SS\ C transition is first order, as can be seen from the
hysteresis in the plot of density as a function of $\tilde{T}$ in Fig. 3(a).
The region of metastability associated with this transition is in the range $%
\tilde{T}_{c}(=0.043)<\tilde{T}<\tilde{T}_{c_{o}}$. The SS\ C phase
continues to higher temperature, up to point $\mathbf{b}$. The SS\ C to
solid transition point $\mathbf{b}$ (of BKT type) is indicated by the second
arrow in Fig. 3(a). The area delimited by $\mathbf{c0-c1-b}$ contains the
ferrimagnetic $(x,-x^{\prime },x^{\prime \prime })$ supersolid phase that
evolves from SS C with increasing $\widetilde{\mu }$ for $\tilde{T}_{c_{o}}<%
\tilde{T}<\tilde{T}_{b}$. The $\mathbf{c0-c1}$ line is second order.
Dependence on $\widetilde{\mu }$ at fixed $\tilde{T}=0.06$ for the
superfluid order parameter and the density is shown in Fig. 3(b). The arrow
to the right indicates the BKT transition from SS\ A to S: In Ref. \cite%
{Boninsegni}, a BKT transition from SS A to S was also found with QMC at
very similar temperatures. The arrow to the left marks the transition from
SS\ C to SS\ A. The region $\mathbf{b-c1-c2}$ delineates the solid order
that evolves from a $(x,-x,0)$ pattern. Outside this region, the solid phase
has ordering $(x,x,-x^{\prime })$ and the transition between the two types
of solids is second order.

%      We note that at h=0 temperature drives
%      SS A to SS C  with a first order transition. The
%      point  'a'  is a spiondal point $T_SSA=0.053$ below this
%      point SSA state can be found and above other spinodal point
%      $T_{\it{SSC}}=0.0430$, which is not shown in the digram, {\it{SSC}}
%      state can be found. Area  'abc' indicate regions of  supersolid which
%      evolves from SS C with increasing $h$ at fixed value of $T/V$.
%      In Fig. 4(a) and (b), we show density  (magentization) hysteresis and average valu
%of $<S_{x}>$ as a function of $T/V$ at h=0 and  as a function of $h$ at $T/V=0.06$
%       respectively. We can notice that two kinks one indicates
%       a second order transition from SS C and SS A  and other
%       the KT transition from  SS A to S.

\begin{figure}[tbp]
\includegraphics[scale=0.3]{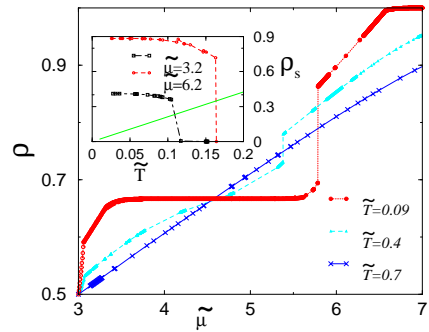}
\caption{Density as a function of $\widetilde{\protect\mu }$ for various
values of $\tilde{T},$ each of which corresponds to a different horizontal
cut on the phase diagram of Fig. 2. The inset shows the BKT transition in $%
\protect\rho _{s}$ as a function of $\tilde{T}$ at $\widetilde{\protect\mu }%
=3.2$ and $6.2$. The straight diagonal line is the BKT prediction. }
\label{fig:finitetd}
\end{figure}

The plot of density as a function of $\widetilde{\mu }$ in Fig.\ 4 confirms
the order of the last two transitions we mentioned. The first kink in the $%
\tilde{T}=0.4$ and that in the $\tilde{T}=0.1$ curves are associated with,
respectively, the second order solid to solid and SS C to SS A (at finite $%
\widetilde{\mu }$ where both phases are ferrimagnetic). %\begin{figure}
%\includegraphics[scale=0.4]{rho_Sx_vs_temp_h0_vs_h_temp0.06.eps}
%\includegraphics[scale=0.4]{phased_temp_rho_plane_at_Delta_0.2.eps}
%\caption{Phase diagram at $\Delta=0.2$ as a functions of (a)h and T and (b) $\rho$ and T. %Several
%phases appear: SF, SS A, SS C, S, normalfluid(NF), full, and phase seperated(PS).
%The C indicates SS C.}
%\label{fig:Finitetph}
%\end{figure}

\textit{Finite t'. }Finally, we investigate whether second-neighbor hopping $%
t^{\prime },$ in the particle-hole symmetry case $\widetilde{\mu }=3,$ can
induce the SS\ C phase at zero temperature, as proposed in Ref. \cite{DVT}.
A finite $t^{\prime }$ allows same sublattice hopping. In the presence of $%
t^{\prime }$, we choose clusters $`$1' and $`$2' shown in the inset of
Fig.1, and connect them to each other through the perturbation $t^{\prime }$%
. The effect of the other bonds that connect clusters '1' and '2' are
included in the self-consistent Eqs. (\ref{eq:selfc_field}). We checked that
the ground state energy of this cluster is lower than that of cluster '3'
(where all bonds reside on the cluster).
\begin{figure}[tbp]
\includegraphics[scale=0.35]{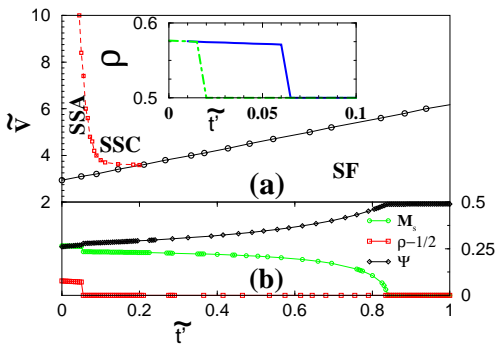}
\caption{(a) The ground-state phase diagram for the $t-t^{\prime }-V$ model
at $\widetilde{\protect\mu }=3^{+}$. The solid line indicates second order
transiton, whereas broken line indicates the first odrer transition. $%
t^{\prime }$ and V are measured with respect to t. The inset shows $\protect%
\rho $ vs $t^{\prime }$ hysteresis curve at $\widetilde{V}=6$. (b) $M_{s}$, $%
\protect\rho -1/2$, and $\Psi $ as a function of $t^{\prime }$ for $%
\widetilde{V}=6.$}
\label{fig:dia_h_0_V_vs_tp}
\end{figure}

The ground-state phase diagram for $\widetilde{\mu }=3^{+}$ is shown in Fig. %
\ref{fig:dia_h_0_V_vs_tp} (a). We note that for $\widetilde{V}=V/t>3.0$ a
small value of the perturbation $\widetilde{t^{\prime }}=t^{\prime }/t$
drives SS\ A to SS\ C through a strong first order transition, as can be
seen from the hysteresis exhibited in the inset of Fig.\ \ref%
{fig:dia_h_0_V_vs_tp} (a). In part (b) of the same figure, we plot the
staggered density $M_{s}$, the superfluid order parameter $\Psi $, and the
average value of $\rho -1/2$ as a function of $\widetilde{t^{\prime }}$ at $%
\widetilde{\mu }=3,$ corresponding to an horizontal cut at $\widetilde{V}=6$
in the phase diagram. We note that the finite value of $\rho -1/2$
corresponds to SS A. With increasing $\widetilde{t^{\prime }}$ the value of $%
\rho -1/2$ jumps to zero, indicating SS C. Similar jumps can be seen in the
other two curves.

%{\bf {In ref\cite{DVT} it was suggested as candidate for $NCCP^{1}$(it can
%found in the second last paragraph before Acknowlegment section)}}\\

In summary, the strong geometric frustration present on the triangular
lattice has striking consequences on the phase diagram of hard core bosons.
First, as is well known, it allows the \textquotedblleft
ferrimagnetic\textquotedblright\ supersolids SS A and SS B phases to appear
at $T=0$. Second, the triangular lattice is associated with strong entropic
effects at finite $T$ that, we have shown, lead to a pronounced Pomeranchuk
effect. We have also shown at the particle-hole symmetric point $\widetilde{%
\mu }=3$ that entropic effects at finite $T,$ or finite $t^{\prime }$ at $%
T=0,$ lead to the appearance of the elusive \textquotedblleft
antiferromagnetic\textquotedblright\ SS\ C phase. Since the SS\ A and\ SS\ B
supersolids break particle-hole symmetry, it is natural that increasing
temperature restores a symmetric SS\ C state. In the case of $t^{\prime },$
it is a simple exercise to show that for same sublattice hopping, kinetic
energy is minimized by $\left( \left\vert 0\right\rangle +\left\vert
1\right\rangle \right) /\sqrt{2},$ i.e. the $0$ state in spin language.
Finite $t^{\prime }$ thus also favours the restoration of the SS\ C\ $%
(x,-x,0)$ state. The SS\ A to SS\ C transition is strongly first order under
the influence of either $T$ or $t^{\prime }$ at $\widetilde{\mu }=3^{+}$. It
is clearly not possible to see non-Landau Quantum Critical Point \cite%
{Senthil} with SCMFT, nevertheless it is likely that transitions that are
strongly first order in SCMFT will not become continuous unless quantum
fluctuations beyond the cluster size are singular enough to completely drive
the transition. This is a delicate point that requires much more detailed
studies guided by our results for phase boundaries. Our finite temperature
results are important for experimental studies of this very rich phase
diagram with optical lattices or in solid state XXZ spin analogs. 
%   First order transition from SS A
%  to SSC has ruled out exotic non-LGW deconfined qunatum criticality. 

SRH thanks A.Georges and R.Moesnner for stimulating discussions at the
initial stage of the project. We also thank M. Boninsegni, B. Davoudi, B.
Kyung, A.H. Nevidomskyy, A. Paramekanti and N. Prokof'ev for useful
conversations. Computations were performed on the Ms RQCHP cluster. The
present work was supported by NSERC (Canada), CFI (Canada), CIAR, the Tier I
Canada Research Chair Program (A.-M.S.T.) and the Center for Materials
Theory, Rutgers University (LdM).

\end{document}